# ОБ ОСОБЕННОСТЯХ РАСПРЕДЕЛЕНИЙ ПЕРЕФОКУСИРОВАННЫХ РАСПЫЛЕННЫХ АТОМОВ, ЭМИТИРОВАННЫХ С ГРАНИ (001) Ni, ПО УГЛАМ И ЭНЕРГИИ

© 2015 г.   В. Н. Самойлов, А. И. Мусин, Н. Г. Ананьева

*Физический факультет Московского государственного университета имени М. В. Ломоносова, Москва, Россия*

E-mail: *samoilov@polly.phys.msu.ru*

В работе с помощью компьютерного моделирования методом молекулярной динамики исследованы особенности перефокусировки атомов, распыленных с поверхности граней (001) Ni. Обнаружена многозначность сигнала перефокусированных атомов по начальному азимутальному углу вылета $\varphi_0$, связанная с различными механизмами рассеяния атомов. Перефокусированные атомы образуют отдельный максимум и могут быть выделены в экспериментах с разрешением по углам и энергии отдельно от фокусированных и "собственных" атомов.

## ВВЕДЕНИЕ

Анизотропия двумерного углового распределения атомов, распыленных с поверхности низкоиндексных граней монокристалла под действием ионной бомбардировки, является одним из сложных эффектов, отражающих анизотропию структуры поверхности кристаллов. Картина углового распределения распыленных атомов чувствительна к типу облучаемой ионами грани кристалла [1–3]. В расчетах эмиссии атомов с поверхности граней (001) Ni и (111) Ni, в частности, с разрешением по энергии, наблюдались максимумы эмиссии, которые по своей угловой ширине и направлениям формирования соответствовали экспериментально наблюдаемым максимумам эмиссии – пятнам Венера [4]. Таким образом, формирование экспериментально наблюдаемых пятен Венера в двумерном угловом распределении атомов, распыленных с поверхности монокристалла, оказалось возможным объяснить действием только по-верхностного механизма фокусировки. На стадии эмиссии происходит сильное перераспределение вылетающих атомов по углам и энергии, такое, что, стадия эмиссии играет важную роль в формировании углового и энергетического распределений распыленных атомов.

В [4–7] был обнаружен и исследован немонотонный сдвиг максимума полярного углового распределения распыленных атомов с ростом их энергии. В основе эффекта лежит конкуренция двух факторов: блокировки эмитируемых атомов в сторону нормали к поверхности в процессе вылета и преломления на плоском потенциальном барьере. Такой сдвиг максимума также наблюдался экспериментально [8]. Было показано, что основные особенности наблюдаемых угловых распределений с разрешением по энергии описываются взаимодействием эмитируемых атомов с линзами, состоящими из двух атомов – ближайших к вылетающему атому соседей в плоскости поверхности [5, 7].

Основное внимание при исследовании фокусировки распыленных атомов уделялось исследованиям механизмов фокусировки по полярному углу вылета. Исследованиям фокусировки распыленных атомов по азимутальному углу посвящено достаточно небольшое число работ. В этой ситуации нам показалось важным исследовать механизмы формирования анизотропии азимутального распределения распыленных атомов.

Для несимметричных относительно направления <010> интервалов азимутального угла φ формирование сигнала распыленных атомов при эмиссии атомов с поверхности происходит за счет "собственных" атомов, начальный угол вылета которых $φ_0$ принадлежит интервалу углов φ, и фокусировки "несобственных" атомов: фокусированных атомов, рассеянных на ближайшем атоме линзы из двух ближайших к эмитируемому атому атомов поверхности, и перефокусированных атомов, рассеянных на дальнем атоме линзы. Для фокусированных атомов угол φ и угол $φ_0$ лежат по одну сторону от направления <010> на центр линзы из двух ближайших к эмитируемому атому атомов поверхности, для перефокусированных атомов – по разные стороны от этого направления. Таким образом, фокусировка атомов идет к центру линзы из двух атомов, а перефокусировка – через центр линзы из двух атомов поверхности. Эффект перефокусировки был обнаружен в [5, 9] и исследован в ряде работ, например, в [10]. В настоящей работе исследованы особенности фокусировки атомов, рас-пыленных с поверхности грани (001) Ni, по азимутальному углу φ с разрешением по полярному углу и энергии. Ставилась задача изучить вклад перефокусированных атомов в формирование распределений распыленных атомов по углам и энергии. Также изучался вопрос о выделении перефокусированных атомов в общем сигнале эмитированных атомов.

## МОДЕЛЬ РАСЧЕТА

Расчеты были проведены для эмиссии атомов с поверхности грани (001) Ni. Было проведено сравнение результатов моделирования, полученных в рамках двух моделей расчета. В модели I поверхность кристалла моделировалась 20 атомами поверхности, ближайшими к узлу решетки, из которого происходила эмиссия атома (модель 21 атома). Подобная модель использовалась в нашей работе [11]. В модели II поверхность кристалла была представлена минимальным фрагментом – кольцом из четырех атомов поверхности, ближайших к узлу решетки, из которого происходила эмиссия атома (модель 5 атомов). Эта модель использовалась в ряде наших предыдущих работ, в частности, в работе [10].

Для расчета эмиссии атомов использовался метод молекулярной динамики. Взаимодействие эмитируемого атома с атомами поверхности в модели описывалось потенциалом отталкивания, а на достаточно большом удалении атома от поверхности был введен плоский потенциальный барьер. В качестве потенциала взаимодействия атом–атом был использован потенциал Борна–Майера:

$$U(r) = A\exp(-r/b) \qquad (1)$$

с параметрами $A = 23853.96$ эВ и $b = 0.196$ Å для взаимодействия двух атомов Ni из работы [12]. Энергия связи составляла 4.435 эВ.

Атом выбивался из узла на поверхности с энергией $E_0$ под углами $\vartheta_0$ (начальный полярный угол, отсчитывался от нормали к поверхности) и $\varphi_0$ (начальный азимутальный угол, $\varphi_0 = 90°$ соответствовал направлению <010> на центр линзы из двух ближайших к эмитируемому атому атомов поверхности). Начальная энергия $E_0$ изменялась от 0.5 эВ до 100 эВ. Шаг по $E_0$ составлял 0.01 эВ. Шаг по $\varphi_0$ был равен 0.5°, шаг по $1 - \cos\vartheta_0$ составлял 1/450. Было использовано начальное распределение эми-тируемых атомов по углам и энергии $\cos\vartheta_0/E_0^2$ [13, 14]. Таким образом, распределение эмитируемых атомов по начальному азимутальному углу $\varphi_0$ было изотропным.

Считалось, что распыление происходит только за счет атомов поверхностного слоя. Это допущение вполне оправдано ввиду того, что для мишеней, состоящих из средних по массе и тяжелых атомов, вклад атомов поверхностного слоя в распыление является доминирующим (88,6 % для случая ионной бомбардировки Cu [15], 82% для случая ионной бомбардировки Mo [16]). Обсуждение некоторых особенностей и корректности модели, используемой в настоящей работе, приведено также в работе [5].

При выполнении расчетов исследован вопрос, каким образом взаимодействие эмитируемых атомов с атомами поверхности кристалла в процессе вылета влияет на особенности наблюдаемого углового распределения распыленных атомов с разрешением по энергии.

Расчеты были выполнены с использованием ресурсов суперкомпьютерного комплекса МГУ "Ломоносов" [17, 18]. Для этих расчетов программа была написана на языке Fortran 90 с использованием Intel MPI.

## РЕЗУЛЬТАТЫ И ИХ ОБСУЖДЕНИЕ

### *Многозначность распределения перефокусированных атомов по углу $\varphi_0$*

Исследуем более подробно механизмы формирования распределения эмитированных атомов по углам с разрешением по энергии. Для этого рассчитаем, под какими начальными азимутальными углами вылетали с поверхности распыленные атомы, наблюдаемые в фиксированных интервалах углов $\vartheta$ и $\varphi$ и энергии $E$. Заметим, что плоский потенциальный барьер не изменяет азимутальный угол, под которым движется эмитируемый атом после рассеяния на одном или нескольких атомах поверхности (для распыленного атома).

В дифференциальных распределениях распыленных атомов по углам и энергии была обнаружена многозначность сигналов фокусированных и перефокусированных атомов, по углу вылета $\varphi_0$. На рис. 1 представлено распределение по начальному азимутальному углу $\varphi_0$ и энергии $E$ распыленных атомов, эмитированных с поверхности грани (001) Ni, для полярных углов вылета $\vartheta$ [56.3º, 57.8º] и азимутальных углов $\varphi$ [82.5º, 85.5º]. Число распыленных атомов дано в логарифмическом масштабе (шкала справа). Оказалось, что перефокусированные атомы очень чувствительны к выбору модели, чего нельзя сказать о фокусированных атомах. В модели 21 атома высокоэнергетическая часть распределения перефокусированных атомов, наблюдаемая в модели 5 атомов, загибается в сторону низких значений энергии $E$ и углов $\varphi_0$, более близких к направлению на центр линзы. Таким образом, появляется многозначность в распределении перефокусированных атомов по углу $\varphi_0$ и возникает область нулевого сигнала, размер которой увеличивается при уменьшении энергии $E$. В детекторе с узкой угловой апертурой при энергии $E \leq 6$ эВ наблюдаются перефокусированные атомы, эмитированные под разными углами $\varphi_0$. Показано, что эта многозначность связана с двумя различными механизмами рассеяния перефокусированных атомов для различных углов $\varphi_0$. Для перефокусировки атомов верхней ветви существенным является рассеяние на ближайшем атоме линзы, а для перефокусировки атомов нижней ветви – также рассеяние на атоме, расположенном за линзой. Таким образом, в основе многозначности сигнала перефокусированных атомов по углу $\varphi_0$ лежит многократное рассеяние эмитированного атома на атомах поверхности.

### *Отдельные хребты распределения распыленных атомов по энергии и полярному углу для перефокусированных и фокусированных атомов*

Оказалось, что в распределениях с одновременным разрешением по энергии и полярному углу для фиксированных интервалов углов $\varphi$ отчетливо различаются отдельные хребты – максимумы распределений для фокусированных и перефокусированных атомов (рис. 2). Верхний хребет образован в основном фокусированными атомами, нижний – только перефокусированными атомами. Максимум распределения перефокусированных атомов наблюдается в области энергии и полярных углов, при которых нет вылета других групп атомов. Перефокусированные атомы на 100% формируют наблюдаемый сигнал. Таким образом, в экспериментах с разрешением по углам и энергии оказывается принципиально возможным выделить отдельно сигнал только перефокусированных атомов.

### *О наблюдаемости перефокусированных распыленных атомов*

Распределение распыленных атомов по энергии для фиксированных интервалов полярного и азимутального углов (рис. 3) состоит из вкладов фокусированных

атомов – левый максимум, перефокусированных атомов – правый максимум, и небольшого вклада "собственных" атомов при энергиях $E$ от 0 до 1.2 эВ. При этом высокоэнергетический максимум в распределении эмитированных атомов по энергии с разрешением по полярному и азимутальному углам (см. [19]), наблюдаемый при энергиях $E > 20$ эВ на рис. 2, образуется фокусированными, а не перефокусированными атомами.

Аналогично, распределение распыленных атомов по полярному углу для фиксированных энергии и азимутального угла (рис. 4) состоит из вкладов фокусированных атомов – левый максимум, перефокусированных атомов – правый максимум, и вклада "собственных" атомов вблизи нормали к поверхности, при $1 - \cos\vartheta$ от 0 до 12/45. Распределение распыленных атомов по полярному углу для того же интервала азимутальных углов, без разрешения по энергии имеет один максимум, образованный фокусированными и перефокусированными атомами.

## ВЫВОДЫ

С помощью модели молекулярной динамики исследованы особенности перефокусировки атомов, эмитированных с поверхности грани (001) Ni, по азимутальному углу при формировании распределений распыленных атомов с разрешением одновременно по полярному углу и энергии. Исследованы механизмы формирования особенностей этих распределений.

В направлении узел – центр линзы из двух ближайших атомов поверхности различие распределений, рассчитанных по моделям 21 и 5 атомов, связано с рассеянием эмитируемых атомов на атоме, расподоженном за линзой, который присутствует только в модели 21 атома. В модели 21 атома перефокусированные атомы могут образоваться ближе к центру линзы. Этот эффект обусловлен рассеянием эмитированных атомов последующими атомами за линзой.

Рассчитаны дифференциальные распределения распыленных атомов по начальному углу $\varphi_0$ и энергии $E$. Обнаружена многозначность сигналов фокусированных и перефокусированных атомов по углу вылета $\varphi_0$ в модели 21 атома при сравнительно небольших значениях энергии $E$. Показано, что эта многозначность связана с двумя различными механизмами рассеяния перефокусированных атомов для различных углов $\varphi_0$.

Выявлены области значений полярного и азимутального углов вылета $\vartheta$ и $\varphi$ и энергии $E$, для которых сигнал распыленных атомов на 100% формируется за счет эмитированных атомов, перефокусированных относительно центра линзы.

Обнаружено, что в распределениях с одновременным разрешением по энергии и полярному углу для фиксированных интервалов углов $\varphi$ отчетливо различаются отдельные хребты – максимумы распределений для фокусированных и перефокусированных атомов. Показано, что в экспериментах с разрешением по углам и энергии оказывается принципиально возможным выделить отдельно сигнал только перефокусированных распыленных атомов.

# THE PECULIARITIES OF DISTRIBUTIONS OF OVERFOCUSED SPUTTERED ATOMS EJECTED FROM (001) Ni WITH ENERGY AND ANGULAR RESOLUTION

V. N. Samoilov, A. I. Musin, N. G. Ananieva

*Faculty of Physics, M. V. Lomonosov Moscow State University, 119991 Moscow, Russia*

The features of the azimuthal-angle overfocusing of atoms sputtered from the surface of the Ni (001) face are studied by molecular dynamics computer simulation. The signal of overfocused atoms is found to be multi-valued with respect to the initial azimuthal angle $\varphi_0$ due to different mechanisms of scattering. The overfocused atoms form separate maximum and can be detected in experiments with angle and energy resolution separately from the focused and the "own" atoms.

ПОДПИСИ К РИСУНКАМ

**Рис. 1.** Распределение по начальному азимутальному углу $\varphi_0$ и энергии $E$ распыленных атомов для полярных углов вылета $\vartheta$ [56.3°, 57.8°] и азимутальных углов $\varphi$ [82.5°, 85.5°]. В нижней части рисунка – фокусированные атомы, в верхней части – перефокусированные. В отличие от модели 5 атомов отсутствуют перефокусированные атомы с большими энергиями.

**Рис. 2.** Распределение распыленных атомов при эмиссии с грани (001) Ni одновременно по 1 – $\cos\vartheta$ и энергии $E$ для интервала азимутальных углов $\varphi$ [76.5°, 79.5°]. Верхний хребет образован в основном фокусированными атомами, нижний – только перефокусированными атомами.

**Рис. 3.** Распределения всех распыленных атомов (*а*) и только перефокусированных распыленных атомов (*б*) по энергии $E$ при эмиссии с грани (001) Ni для полярных углов вылета $\vartheta$ [56.3°, 57.8°] и интервала азимутальных углов $\varphi$ [76.5°, 79.5°].

**Рис. 4.** Распределения всех распыленных атомов (*а*) и только перефокусированных распыленных атомов (*б*) по 1 – $\cos\vartheta$ при эмиссии с грани (001) Ni для энергии $E$ 3.0 ± 0.1 эВ и интервала азимутальных углов $\varphi$ [76.5°, 79.5°].

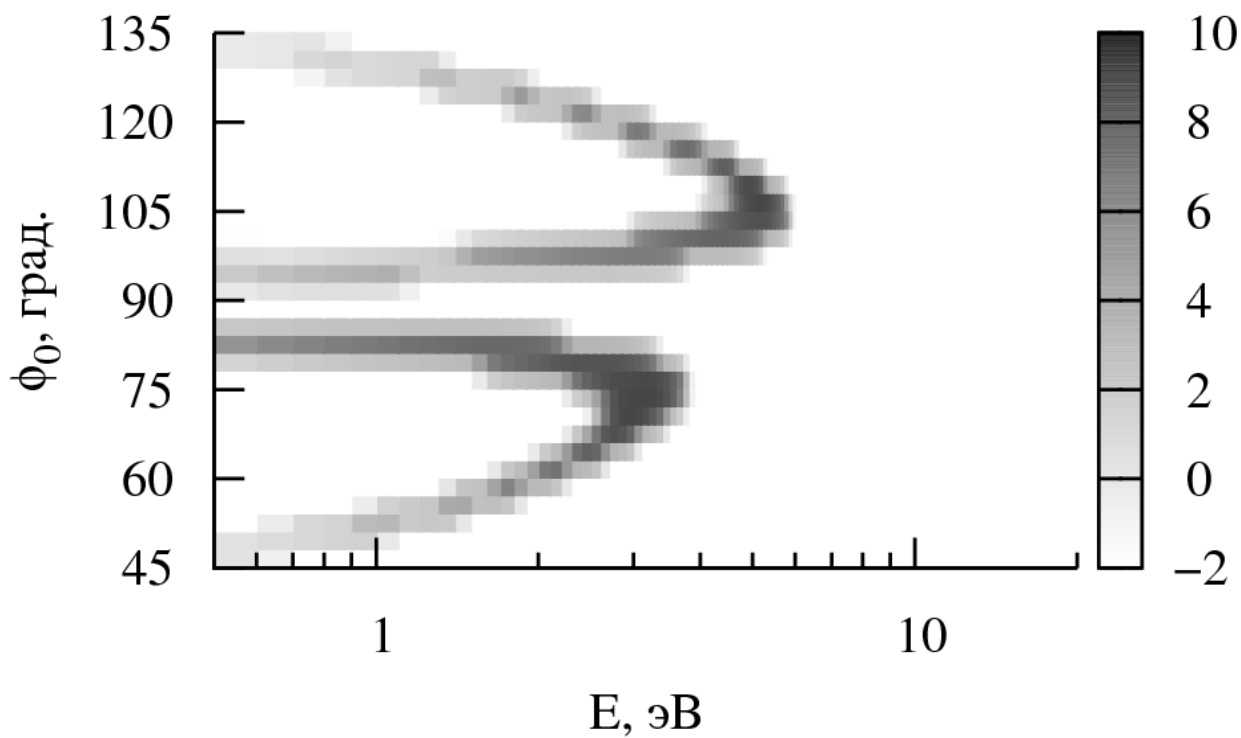

**Рис. 1.** Распределение по начальному азимутальному углу φ₀ и энергии *E* распыленных атомов для полярных углов вылета ϑ [56.3°, 57.8°] и азимутальных углов φ [82.5°, 85.5°]. В нижней части рисунка – фокусированные атомы, в верхней части – перефокусированные. В отличие от модели 5 атомов отсутствуют перефокусированные атомы с большими энергиями.

В. Н. Самойлов, А. И. Мусин, Н. Г. Ананьева
ОБ ОСОБЕННОСТЯХ РАСПРЕДЕЛЕНИЙ ПЕРЕФОКУСИРОВАННЫХ РАСПЫЛЕННЫХ АТОМОВ, ЭМИТИРОВАННЫХ С ГРАНИ (001) Ni, ПО УГЛАМ И ЭНЕРГИИ

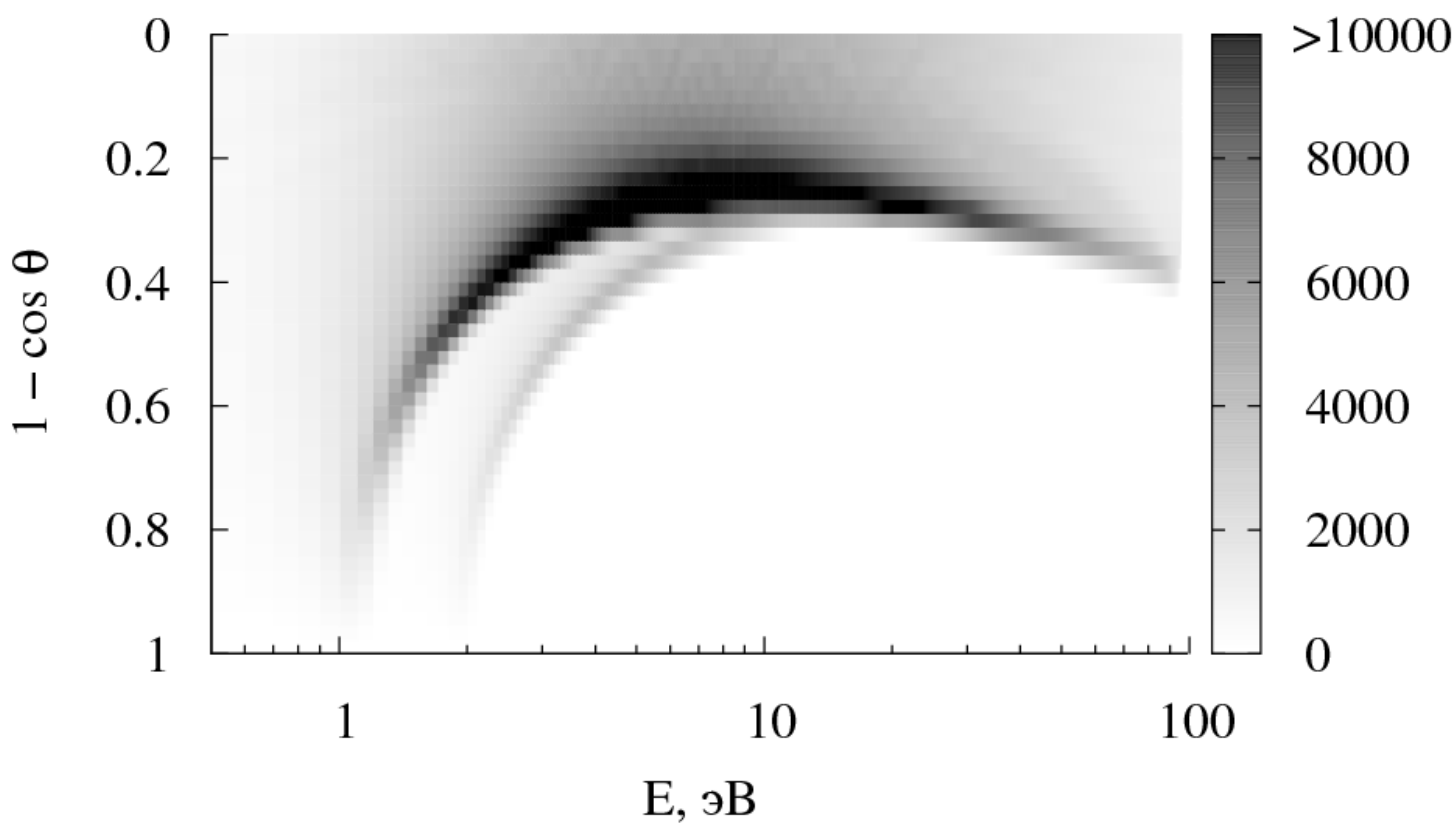

**Рис. 2.** Распределение распыленных атомов при эмиссии с грани (001) Ni одновременно по 1 − cosϑ и энергии $E$ для интервала азимутальных углов φ [76.5°, 79.5°]. Верхний хребет образован в основном фокусированными атомами, нижний – только перефокусированными атомами.

В. Н. Самойлов, А. И. Мусин, Н. Г. Ананьева

ОБ ОСОБЕННОСТЯХ РАСПРЕДЕЛЕНИЙ ПЕРЕФОКУСИРОВАННЫХ РАСПЫЛЕННЫХ АТОМОВ, ЭМИТИРОВАННЫХ С ГРАНИ (001) Ni, ПО УГЛАМ И ЭНЕРГИИ

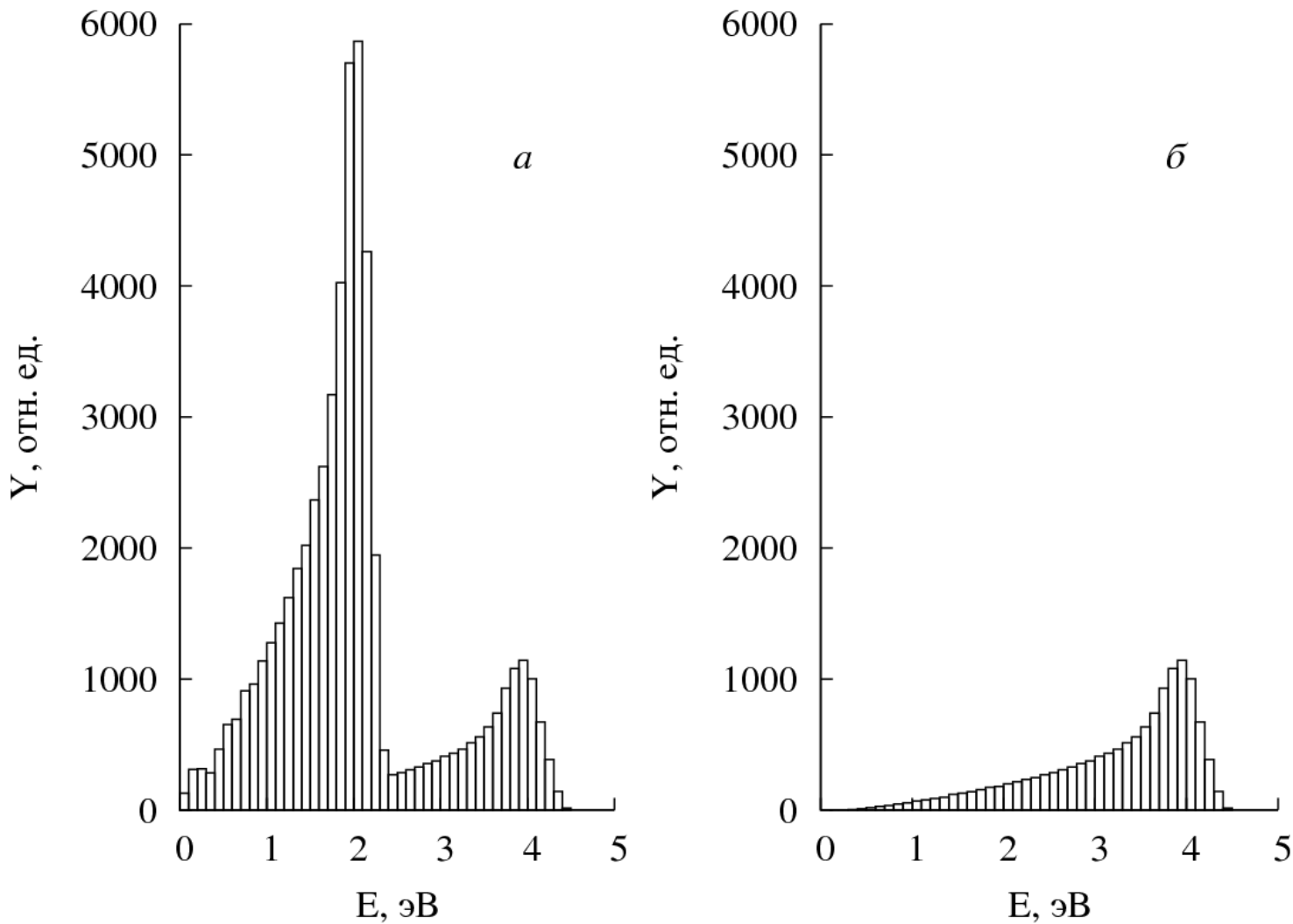

**Рис. 3.** Распределения всех распыленных атомов (*а*) и только перефокусированных распыленных атомов (*б*) по энергии *E* при эмиссии с грани (001) Ni для полярных углов вылета $\vartheta$ [56.3°, 57.8°] и интервала азимутальных углов φ [76.5°, 79.5°].

В. Н. Самойлов, А. И. Мусин, Н. Г. Ананьева
ОБ ОСОБЕННОСТЯХ РАСПРЕДЕЛЕНИЙ ПЕРЕФОКУСИРОВАННЫХ РАСПЫЛЕННЫХ АТОМОВ, ЭМИТИРОВАННЫХ С ГРАНИ (001) Ni, ПО УГЛАМ И ЭНЕРГИИ

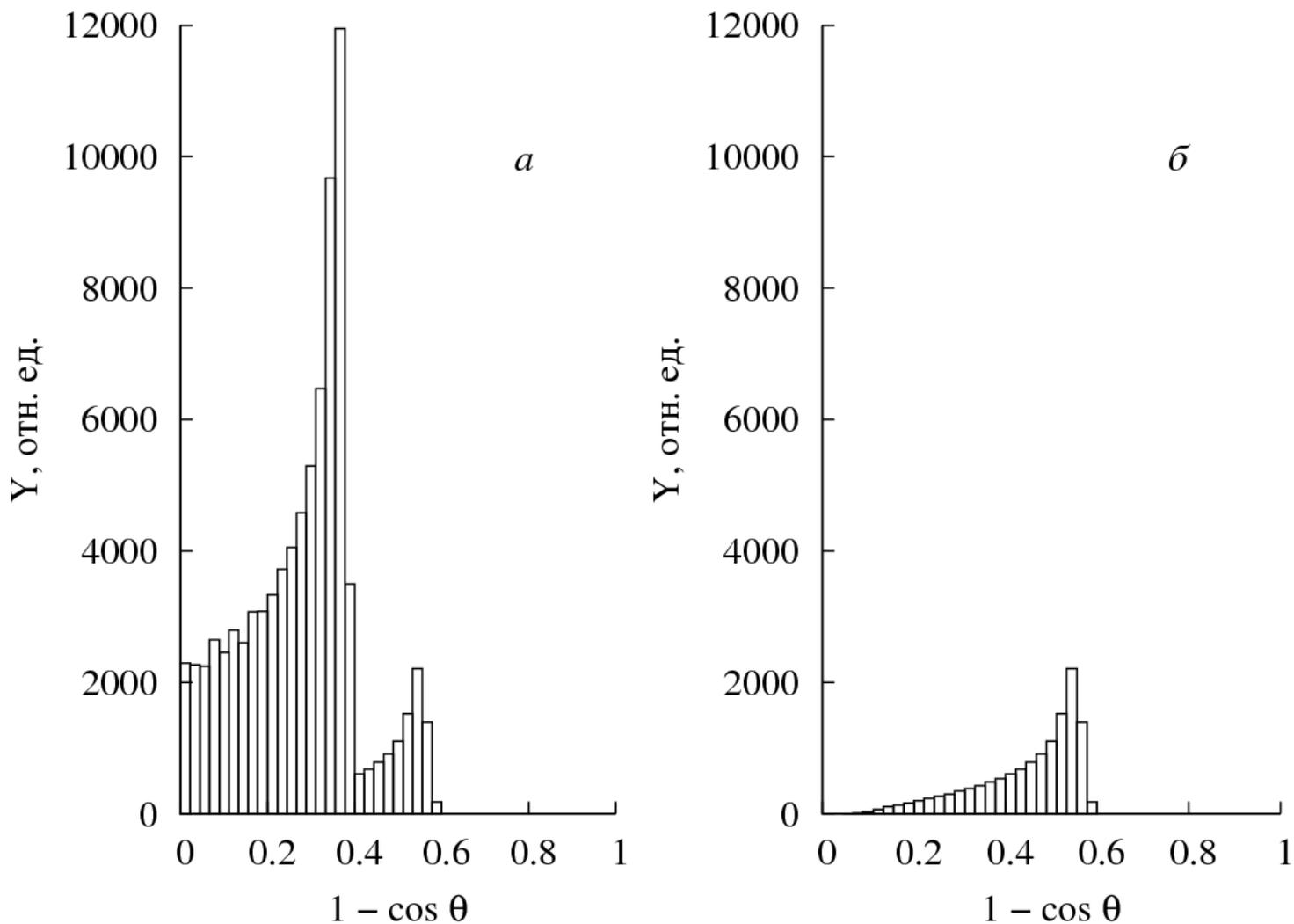

**Рис. 4.** Распределения всех распыленных атомов (*а*) и только перефокусированных распыленных атомов (*б*) по $1 - \cos\vartheta$ при эмиссии с грани (001) Ni для энергии $E$ 3.0 ± 0.1 эВ и интервала азимутальных углов φ [76.5°, 79.5°].

В. Н. Самойлов, А. И. Мусин, Н. Г. Ананьева
ОБ ОСОБЕННОСТЯХ РАСПРЕДЕЛЕНИЙ ПЕРЕФОКУСИРОВАННЫХ РАСПЫЛЕННЫХ АТОМОВ, ЭМИТИРОВАННЫХ С ГРАНИ (001) Ni, ПО УГЛАМ И ЭНЕРГИИ